\documentclass[aip,jap,preprint,groupedaddress,superscriptaddress]{revtex4}

\usepackage{graphicx,epstopdf, amsmath,color}

\usepackage[breaklinks, colorlinks=true,citecolor={blue}, linkcolor= {blue}, urlcolor ={blue}]{hyperref}

\begin{document}

\title{Significant Enhancement of Carrier Mobility in Finite vs. Infinite Square Quantum Wells: A Comparative Study of GaAs/In$_x$Ga$_{1-x}$As/GaAs Heterostructures}

\author{Truong Van Tuan} \email[]{tuanbo83@gmail.com}
\affiliation{University of Tran Dai Nghia, 189 Nguyen Oanh Str., Go Vap Dist., Ho Chi Minh City, Vietnam}
\author{Nguyen Dung Chinh} \email[]{nguyendungchinh@dtu.edu.vn}
\affiliation{Institute of Fundamental and Applied Sciences, Duy Tan University, 06 Tran Nhat Duat St., District 1, Ho Chi Minh City 70000, Viet Nam}
\affiliation{Faculty of Environmental and Natural Sciences, Duy Tan University, 03 Quang Trung St., Hai Chau, Da Nang 50000, Viet Nam}

\author{Tran Trong Tai} 
\affiliation{University of Tran Dai Nghia, 189 Nguyen Oanh Str., Go Vap Dist., Ho Chi Minh City, Vietnam}

\author{Vo Van Tai}
\affiliation{Laboratory of Applied Physics, Science and Technology Advanced Institute, Van Lang University, Ho Chi Minh City, Vietnam} \affiliation{Faculty of Applied Technology, School of Technology, Van Lang University, Ho Chi Minh City, Vietnam}

\author{Nguyen Duy Vy}\email[Corresponding author email: ]{nguyenduyvy@vlu.edu.vn} \affiliation{Laboratory of Applied Physics, Science and Technology Advanced Institute, Van Lang University, Ho Chi Minh City, Vietnam} \affiliation{Faculty of Applied Technology, School of Technology, Van Lang University, Ho Chi Minh City, Vietnam}

\date{\today}

\begin{abstract}
The geometry of quantum wells (QWs) critically influences carrier mobility, yet systematic comparisons between finite and infinite square QWs remain scarce. We present a comprehensive study of GaAs/In$_x$Ga$_{1-x}$As/GaAs heterostructures using a variational-subband-wave-function model, analyzing key scattering mechanisms: remote impurities (RI), alloy disorder (AD), surface roughness (SR), acoustic (ac) and piezoelectric (PE) phonons, and longitudinal optical (LO) phonons. The mobility ratio $R=\mu_{fin}/\mu_{inf}$ reveals distinct trends: $R_{RI}$ and $R_{LO}<$ 1 (long-range Coulomb/inelastic scattering), while $R_{AD}$, $R_{ac}$, $R_{PE}$, $R_{SR}>$ 1 (static potentials). Finite QWs achieve higher mobility at low temperatures (77 K), narrow widths ($<$ 100 \AA ), and low densities, enhanced by high indium content. Conversely, infinite QWs outperform at 300 K due to dominant LO scattering. These findings provide actionable guidelines for optimizing QW-based devices such as HEMTs and lasers across operational regimes.
\end{abstract}

\maketitle 

\section{Introduction}\label{sec:intro}
Semiconductor heterostructures, particularly 
GaAs/In$_x$Ga$_{1-x}$As/GaAs, have revolutionized modern electronics and optoelectronics due to their tunable bandgaps and superior carrier confinement properties \cite{Arent, Bedoui,Vy2020superlattice}. These systems are pivotal in high-electron-mobility transistors (HEMTs) for low-noise ultra-high-frequency applications and in high-efficiency laser diodes, where precise control over carrier dynamics is critical \cite{Vainberg2013}. Central to their performance is carrier mobility, which governs device speed and efficiency. While prior studies have extensively characterized mobility in finite square quantum wells (QWs) under static scattering mechanisms such as surface roughness (SR) and alloy disorder (AD) \cite{Quang_13, Tuan_16}, a systematic comparison between finite and infinite QWs—accounting for both elastic and inelastic scattering processes—remains underexplored.

Recent advances in low-dimensional systems highlight the profound impact of quantum well geometry on mobility. For instance, finite QWs exhibit stronger carrier localization, altering scattering rates compared to infinite QWs \cite{Nestoklon_16}. However, existing theoretical models often neglect the interplay of multiple scattering mechanisms, such as the dominance of longitudinally polarized optical (LO) phonons at room temperature \cite{Kawamura}. This gap limits the ability to predict optimal QW configurations for specific operational conditions (e.g., low-temperature vs. high-temperature regimes). Furthermore, while comparative studies in AlGaN/GaN systems \cite{Tai_17} have demonstrated geometry-dependent mobility trends, analogous insights for 
GaAs/In$_x$Ga$_{1-x}$As/GaAs QWs are lacking.

In this work, we bridge this gap by rigorously analyzing carrier mobility in finite and infinite square QWs using a variational-subband-wave-function model. We incorporate critical scattering mechanisms: static defects (SR, remote impurities (RI), AD), acoustic phonons (via deformation potential and piezoelectric coupling), and inelastic Longitudinal optical phonons treated iteratively \cite{Kawamura, Hamaguchi}. Our study reveals how the mobility ratio $R=\mu_{fin}/\mu_{inf}$ depends on well width ($L$, indium content ($x$), and temperature ($T$). Key findings include the superior mobility of finite QWs at low $T$ and narrow $L$, driven by suppressed RI and enhanced AD scattering, contrasted by the dominance of infinite QWs at room temperature due to LO phonon scattering. These results provide actionable guidelines for designing QW-based devices, balancing trade-offs between structural parameters and operational environments.

\section{Theory}
We investigate a quasi-2D electron gas (Q2DEG) in the $xy$ plane utilizing a finite-depth quantum well model. At low temperatures, we assume that electrons definitively occupy only the lowest conduction sub-band, as established in reference \cite{Quang_13}
\begin{equation}
\Psi(z) =C\sqrt{\frac{2}{L}} 
\begin{cases}
\cos(\frac{1}{2} k L) $e$^{\kappa z}, & $for $z$ $ <0, \\
\cos[k(z-\frac{1}{2}L)], & $for $ 0\leq z \leq L, \\ 
\cos(\frac{1}{2} k L) $e$^{-\kappa (z-L)}, & $for $z$ $ > L,
\end{cases}
 \end{equation}
The constant $C$ serves as a crucial normalization factor that is established by
\begin{equation}
C^2\left(1+\frac{\sin{a}}{a}+\frac{1+\cos{a}}{b}\right)=1.
 \end{equation}
In this context, $a = kL$ and $b=\kappa L$ represent essential dimensionless quantities based on the thickness of the quantum well (QW), $L$, and the barrier height $V$. Understanding these relationships is crucial for optimizing performance in various applications.
\begin{equation}
 a=\frac{L\sqrt{2 m_z V}}{\hbar}\cos{\frac{1}{2}a},
\end{equation}
and
\begin{equation}
 b=a\tan(a/2),
\end{equation}
with $m_z$ being the effective mass of the electron along the growth direction. 
In the limiting scenario where $V$ approaches infinity, it follows that $C$ equals 1 and $a$ equals $\pi$. This leads us to the wave function for a finite square quantum well, which is articulated in \cite{Tuan_18}. 
The mobility in the Boltzmann formalism, using the relaxation time approximation, is expressed by \cite {Kawamura, Hamaguchi}
\begin{equation}
 \mu=\frac{e}{m^*}\langle\tau_{tot}(E)\rangle,
\end{equation}
where
\begin{equation}
 \langle\tau_{tot}(E)\rangle=\frac{\int_{0}^{+\infty}E\frac{\partial f_{0}(E)}{\partial E}\tau(E)dE}{\int_{0}^{+\infty}Ef_{0}(E)dE}
\end{equation}
where $\tau(E)$ represents the relaxation of the electrons, while $f_{0}(E)$ is the Fermi-Dirac distribution. The total mobility is obtained via the Matthiessen's rule
\begin{equation}
 \frac{1}{\tau_{total}(E)}=\frac{1}{\tau_{def}(E)}+\frac{1}{\tau_{phonon}(E)},
\end{equation}
where $\tau_{def}(E)$ and $\tau_{phonon}(E)$ is the relaxation time from defects and phonons scatterings, respectively.
In Boltzmann's theory, the concept of relaxation time is crucial for understanding how systems return to equilibrium is defined as \cite{Khanh_22, Khanh_23}
\begin{align}
 \frac{1}{\tau(E)} &=\frac{m^*}{\pi\hbar^3}\int_{0}^{\pi}d\theta(1-\cos{\theta})\frac{\langle|U(q)^2|\rangle}{\epsilon^2 (q,T)}, \\
%
 \epsilon(q,T) &=1+\frac{2\pi e^2}{\varepsilon_L}\frac{1}{q}F_{C}(q)
[1-G(q)]\Pi(q,T),\\
%
 \Pi(q,T) &=\frac{\beta}{4}\int_{0}^{\infty}d\xi'\frac{\Pi^{0}(q,\xi')}{\cosh^2\left[{\frac{\beta}{2}(\xi-\xi'})\right]},\\
%
 \Pi^{0}(q,E_F ) &=\Pi^{0}(q)=\frac{m^*}{\pi\hbar^2}\left[1-\sqrt{1-\left(\frac{2 k_F}{q}\right)^2}\Theta(q-2 k_F)\right],\\
%
 F_C(qL) &=\frac{1}{C^4}\int_{-\infty}^{+\infty}dz\int_{-\infty}^{+\infty}dz'|\Psi(z)|^2 |\Psi(z')|^2 e^{-q|z-z'|}.
 \end{align}
In this context, let $\varepsilon_L$ denote the background static dielectric constant, and $\Pi^{0}(q)$ represent the zero-temperature polarizability of the 2DEG within the random phase approximation (RPA). The parameter $\beta$ is defined as $\beta=(k_B T)^{-1}$, where $k_B$ is the Boltzmann constant and $T$ is the temperature. The function $f_{0}(E)$ is given by: $f_{0}(E)=1/\{ {\exp[(E-\xi(T))/(k_BT)]+1}\}$; here $\xi(T)$ is defined as: $\xi(T)=k_BT\ln[-1+\exp(E_F/(k_BT))]$; The Fermi energy $E_F $ is expressed as: $E_F =\hbar^2 k_F ^2 /(2 m^*)$ where $k_F=(2\pi N_s )^{1/2}$, with $N_s$ being the carrier density. The function $G(q)$ represents a local-field correction (LFC) that accounts for exchange-correlation effects beyond the RPA \cite{Tai_24}. Additionally, $\langle|U(q)^2|\rangle$ signifies the random potential, which depends on the specific scattering mechanism involved.
In the context of remote impurities scattering (RI), the characterization of the random potential is thoroughly detailed in reference \cite{Tuan_18}.

In surface roughness scattering (SR), the characterization of random potential plays a crucial role, as indicated in the literature \cite{Quang_13, Tuan_18}.
With the roughness-induced piezoelectric scattering (PE), the random potential has the form \cite{Quang_13, Tuan_25}.
For the 2DEG located on the side of a InGaAs layer, the autocorrelation function for alloy disorder scattering (AD) is supplied in the form \cite{Tuan_25, Quang_26}.
Since acoustic (ac) phonons dissipate negligible energy, the quasi-elastic approximation becomes applicable. This allows us to derive a closed-form expression for the relaxation rate \cite{Kawamura},
\begin{equation}
\begin{split}
\frac{1}{\tau_{j}(E)} =& \frac{m^*}{\hbar^3}\frac{4}{(2\pi)^2}\int_{0}^{\pi}d\theta(1-\cos{\theta)}\int_{0}^{+\infty}dq_{z}\frac{\left|C_{j}(q,q_z)\right|^2}{\epsilon^2 (q,T)}|I(q_z)|^2\times \\
 &\times  \frac{1}{1-f_{0}(E)}\left(n_{Q}[1-f_{0}(E+\hbar\omega_Q)]+(n_{Q}+1)[1-f_{0}(E-\hbar\omega_Q)]\right),
\end{split}
\end{equation}
where $\vec{Q}=(\vec{q},q_z)$ and $\hbar\omega_{\vec{Q}}$ is the wave-vector and the energy of 3D phonons,  $n_Q$ the Bose factor, $q=|\vec{k'}-\vec{k}|$, $q^2 =2 k^2 (1-\cos{\theta})$, and $C_{j}(q,q_z)$ are given by 
\begin{equation}
 |C_{DP}(Q)|^2=\frac{D^2\hbar Q}{2\rho u_l}=\frac{D^2\hbar\sqrt{q^2 +q_{z}^2 }}{2\rho u_l}
\end{equation}
for the deformation potential, $D$ is defined as the deformation potential constant, $\rho$ denotes the mass density, and $u_l$ and $u_t$ represent the longitudinal and transverse sound velocities, respectively.
\begin{subequations}
 \begin{align}
 |C_{PE;l}(Q)|^2&=\frac{(eh_{14})^2}{2\rho u_l}\frac{1}{Q}A_{l}(Q),\\
 |C_{PE;t}(Q)|^2&=\frac{(eh_{14})^2}{2\rho u_t}\frac{1}{Q}A_{t}(Q),
 \end{align}
\end{subequations}
for the longitudinal and transverse PE-coupled scatterings, where
\begin{subequations}
 \begin{align}
 A_{l}(q,q_z)&=\frac{9 q_{z}^2 q_{z}^{4}}{2(q_{z}^2 +q^2)^3}, \\
 A_{t}(q,q_z)&=\frac{8 q_{z}^{4}q_{z}^2 +q^6}{4(q_{z}^2 +q^2)^3}, 
 \end{align}
\end{subequations}
where $I(q_z)$ is the overlap function for the intraband scattering,
\begin{equation}
 I(q_z)=\int_{-\infty}^{+\infty}|\Psi(z)|^2 \exp{(i q_{z}z)}.
\end{equation}
In the realistic model of finitely deep square quantum wells, the expression for $|I(q_z)|^2$ is presented in reference \cite{Tuan_18}. 
The scattering of electrons by polar optical phonons plays a crucial role in shaping the electron transport properties of polar semiconductor systems, particularly at elevated temperatures. Due to the inelastic nature of scattering from polar longitudinal optical (LO) phonons, the relaxation time approximation is not applicable. This necessitates a direct approach to solving the Boltzmann transport equation (BTE). The two-dimensional BTE can be effectively formulated as a difference equation that links $\phi(E)$ to $\phi(E\pm\hbar\omega_0)$ as \cite{Kawamura}
\begin{equation}
 1=S_{0}(E)\phi(E)-S_{a}(E+\hbar\omega_0)\phi(E+\hbar\omega_0)-S_{e}(E-\hbar\omega_0)\phi(E-\hbar\omega_0)
\end{equation}

The term $S_{0}(E)$ represents the total contributions from quasi-elastic processes, which include both in-scattering and out-scattering from acoustic phonons, as well as the out-scattering contribution from longitudinal optical (LO) phonons. Meanwhile, $S_{a}(E)$ and $S_{e}(E)$ denote the in-scattering contributions resulting from inelastic scattering processes involving LO phonons. The expression for $S_{0}(E)$ can be written as follows,
\begin{equation}
 S_{0}(E)=S_{0(ac)}(E)+S_{0(op)}(E)
\end{equation}
with
\begin{subequations}
\begin{align}
S_{0({ac})}(E) &= \frac{1}{\tau_{t({ac})}(E)} = \frac{1}{\pi} \int_0^{\pi} d\theta \, (1 - \cos \theta) \frac{\langle|U_{{ac}}(q) |\rangle^2}{\varepsilon^2(q,T)}, 
\end{align}
\text{and}
\begin{align}
S_{0({op})}(E) &= \frac{m^* \omega_0 e^2}{\hbar^2} \left( \frac{1}{\varepsilon_{L\infty}} - \frac{1}{\varepsilon_L} \right) \frac{1}{1 - f_0(E)} 
\big[ n_0 [1 - f_0(E + \hbar \omega_0)] I_{{LO}}^+(E) \nonumber \\
& + (n_0 + 1)[1 - f_0(E - \hbar \omega_0)]\Theta(E-\hbar\omega_{0}) I_{{LO}}^-(E) \big]. 
\end{align}
\end{subequations}
The terms $S_{a}(E)$ and $S_{e}(E)$ can be expressed in the following clear and explicit forms:
\begin{subequations}
\begin{align}
S_a(E) &= \frac{m^{*}\omega_0 e^2}{\hbar^2} 
\left( \frac{1}{\varepsilon_{L\infty}} - \frac{1}{\varepsilon_L} \right)
\frac{1}{1 - f_0(E)} \nonumber \\
& \times  \left\{ n_0 \left[1 - f_0(E + \hbar \omega_0)\right] 
\sqrt{\frac{E + \hbar \omega_0}{E}} 
\, \theta(E - \hbar \omega_0) I_{\mathrm{LO}}^+(E) \right\}, 
\end{align}
\begin{align}
S_e(E) &= \frac{m^* \omega_0 e^2}{\hbar^2} 
\left( \frac{1}{\varepsilon_{L\infty}} - \frac{1}{\varepsilon_L} \right)
\frac{1}{1 - f_0(E)} \nonumber \\
& \times  \left\{ 
(n_0 + 1) \left[1 - f_0(E - \hbar \omega_0)\right] 
\sqrt{\frac{E - \hbar \omega_0}{E}} \right\}\Theta(E - \hbar \omega_0) J_{\mathrm{LO}}^-(E), 
\end{align}
\end{subequations}
where the angular integrals $I_{\mathrm{LO}}^{\pm}(E), J_{\mathrm{LO}}^{\pm}(E)$ are defined by
\begin{subequations}
\begin{align}
I_{\mathrm{LO}}^{\pm}(E) &= \int_0^{\pi} F_C(q_\pm)\, d\theta, \\
J_{\mathrm{LO}}^{\pm}(E) &= \int_0^{\pi} F_C(q_\pm)\cos\theta\, d\theta, 
\end{align}
\end{subequations}
with
\begin{equation}
q_\pm = \left( \frac{2m^*}{\hbar^2} \left[ 2E \pm \hbar \omega_0 
- 2\sqrt{E(E \pm \hbar \omega_0)}\cos\theta \right] \right)^{1/2}. 
\end{equation}

To solve Eq. (20) efficiently, we employ the Ritz iteration method \cite{Kawamura, Rode}. At low temperatures, the in-scattering terms $S_{a}(E)$ and $S_{e}(E)$ can be neglected, leading to the following low-temperature approximation for the relaxation time.
\begin{equation}
 \tau_{LT}(E)=\frac{1}{S_{0}(E)}
\end{equation}
At high temperatures, the electron energy $E$ substantially exceeds $\hbar\omega_0$. This leads to the conclusion that $\phi(E\pm\hbar\omega_0)$ closely approximates $\phi(E)$. Therefore, we can confidently utilize the high-energy (HE) approximation for our analyses.
\begin{equation}
 \tau_{HE}(E)=\frac{1}{S_{0}(E)-S_{a}(E)-S_{e}(E)}
\end{equation}
In the case of finitely deep square quantum wells, it is crucial to refer to Ref. \onlinecite{Tuan_18} for a clear understanding of the definitions of $S_{a}(E)$, $S_{e}(E)$ and $S_{0}(E)$.

\section{Numerical results}
The barrier height of QW for electrons is the conduction band offset, which depends on the In content as $V=0.6(1144 x - 255 x^2 )$ $\text{meV}$ \cite{Quang_13}. The remaining variables are extrapolated from the formula $A = (1 – x) A_\text{InAs} + x.A_\text{GaAs}$ \cite{Bedoui, Fraj, Quay}. The Table  \ref{table} presents the constants for InAs, GaAs, and the finite square quantum well $\text{GaAs}/\text{In}_{x}\text{Ga}_{1-x}\text{As}/\text{GaAs}$.
\begin{table*}[htbp] \centering
\caption{Material parameters for GaAs, InAs, and finite square quantum wells GaAs/In$_x$Ga$_{1-x}$As/GaAs.}
\begin{tabular}{|l|c|c|c|c|c|}
\hline
\textbf{Material Property} & \textbf{GaAs} & \textbf{InAs} & \textbf{In$_{0.1}$Ga$_{0.9}$As} & \textbf{In$_{0.9}$Ga$_{0.1}$As} & \textbf{Refs.} \\
\hline
Barrier height of QW’s $V$ (meV) & -- & -- & 111.85 & 823.05 & \cite{Quang_13} \\
\hline
Effective mass of charge carriers $m^*/m_e$ & 0.067 & 0.024 & 0.0627 & 0.0283 & \cite{Quang_13,VanTan19PLA} \\
\hline
Effective mass along growth direction $m_z/m_e$ & 0.067 & 0.024 & 0.0627 & 0.0283 & \cite{Quang_13} \\
\hline
Lattice constant $a$ (\AA) & 5.64191 & 6.0584 & 5.68356 & 6.01675 & \cite{Quang_13} \\
\hline
Elastic stiffness constants $c_{11}$ ($10^{10}$ Pa) & 11.88 & 8.0 & 11.492 & 8.388 & \cite{Quang_13} \\
Elastic stiffness constants $c_{12}$ ($10^{10}$ Pa) & 5.38 & 5.1 & 5.352 & 5.128 & \cite{Quang_13} \\
Elastic stiffness constants $c_{44}$ ($10^{10}$ Pa) & 5.94 & 4.05 & 5.751 & 4.239 & \cite{Quang_13} \\
\hline
Background static dielectric constant $\epsilon$ & 13.18 & 14.55 & 13.317 & 14.413 & \cite{Quang_13} \\
\hline
Piezoelectric constant $e_{14}$ (C/m$^2$) & -0.16 & -0.045 & 0.1485 & 0.0565 & \cite{Quang_13} \\
\hline
Mass density $\rho$ (g/cm$^3$) & 5.32 & 5.68 & 5.356 & -- & \cite{Quang_13} \\
\hline
Deformation potential constant $D$ (eV) & 11 & 5.04 & 10.404 & 5.636 & \cite{Quay} \\
\hline
Longitudinal sound velocity $u_l$ (cm/s) & $4.73\times 10^5$ & $3.83\times 10^5$ & $4.64\times 10^5$ & $3.92\times 10^5$ & \cite{Quay} \\
\hline
Transverse sound velocity $u_t$ (cm/s) & $3.35\times 10^5$ & $2.64\times 10^5$ & $3.28\times 10^5$ & $2.71\times 10^5$ & \cite{Quay} \\
\hline
Piezoelectric tensor constant $h_{14}$ (V/cm) & $1.45\times 10^7$ & $0.35\times 10^7$ & $1.34\times 10^7$ & $0.46\times 10^7$ & \cite{Quay} \\
\hline
Optical phonon energy $\hbar\omega$ (meV) & 35 & 30 & 34.5 & 30.5 & \cite{Quay} \\
\hline
Optical dielectric constant $\epsilon$ & 10.89 & 12.3 & 11.03 & 12.16 & \cite{Quay} \\
\hline
Impurity position $z_i$ & -- & -- & $-L$ & $-L$ & \cite{Tuan_18} \\
\hline
Roughness correlation length $\Delta$ (\AA) & -- & -- & 2 & 2 & \cite{Vainberg2013,Tuan_18} \\
\hline
Average roughness height $\Lambda$ (\AA) & -- & -- & 100 & 100 & \cite{Vainberg2013,Tuan_18} \\
\hline
Scattering potential of alloy atoms $u_{al}$ & -- & -- & 0.33 & 0.33 & \cite{Vainberg2013} \\
\hline
Impurity concentration NRI (cm$^{-2}$
)& & & $N_s$& $N_s$ & \cite{Tuan_18} \\
\hline
\end{tabular}
\label{table}
\end{table*}

\begin{figure*}[!htb] \centering
\includegraphics[width=0.98\textwidth]{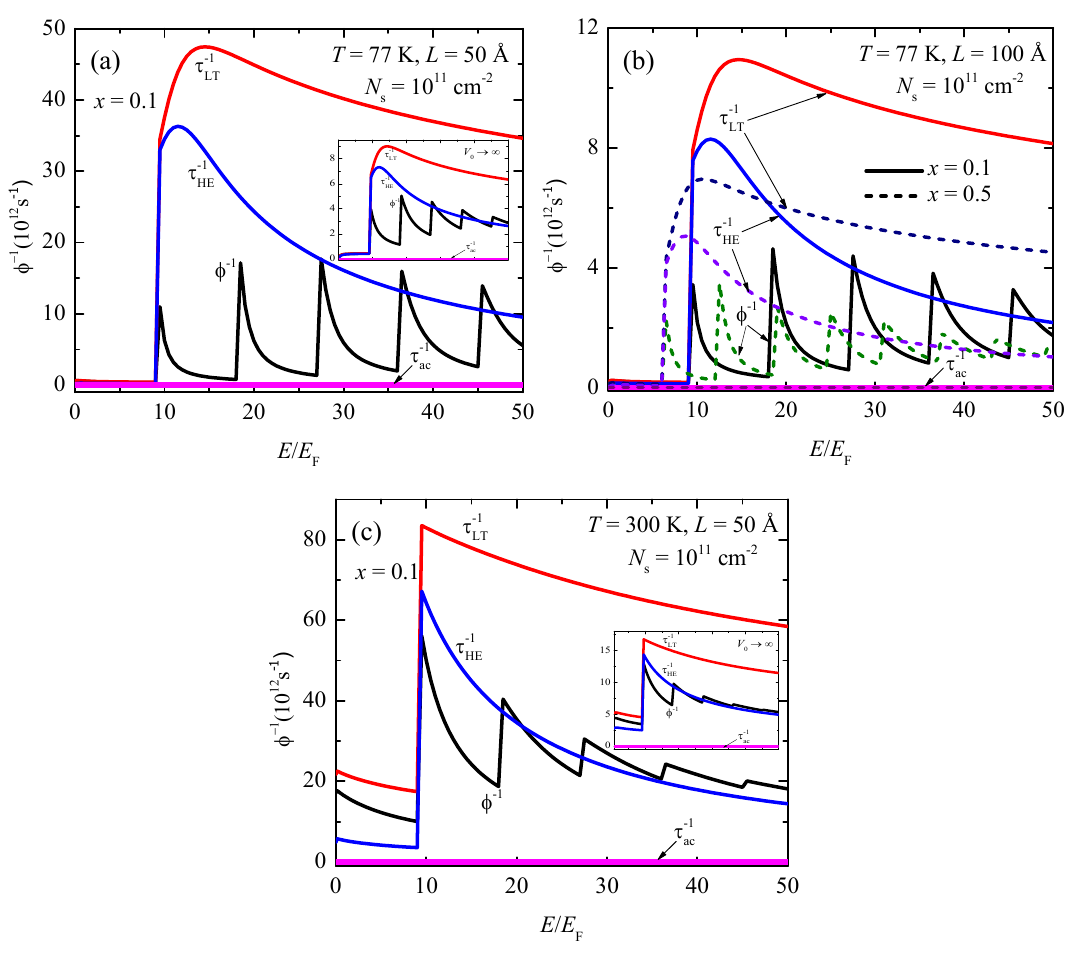}
\caption{Dependence of the perturbation distribution function $\phi^{-1} (E_{\textbf{k}})$ on energy, the acoustic phonon scattering relaxation time $\tau_{ac}^{-1} (E_{\textbf{k}})$, the high-energy relaxation time $\tau_{HE}^{-1} (E_{\textbf{k}})$, and the low-temperature relaxation time $\tau_{LT}^{-1} (E_{\textbf{k}})$ for finite and infinite square quantum wells of $\text{GaAs}/\text{In}_{x}\text{Ga}_{1-x}\text{As}/\text{GaAs}$ with $N_{s}$ = 1$\times  10^{11}$ 
 $\mathrm{cm}^{-2}$ : (a) $L$ = 50 $\mathrm{\AA}$, $T$ = 77 K for $x$ = 0.1 ($V$ = 111.85 meV) and $V\to\infty$; (b) $L$ = 100 $\mathrm{\AA}$, $T$ = 77 K for $x$ = 0.1 and $x$ = 0.5 ($V$ = 508.25 meV) and (c) $L$ = 50 $\mathrm{\AA}$, $T$ = 300 K for $x$ = 0.1 and $V\to\infty$.}
 \label{fig1}
\end{figure*}

Longitudinally polarized optical phonon inelastic scattering (LO) is crucial in the behavior of semiconductor materials at room temperature ($T$ = 300 K). The effectiveness of an iterative method for approximating this phenomenon has been validated in numerous studies \cite{Tai_17, Kawamura}, highlighting its significance in advancing our understanding of semiconductor physics. 

In Fig. \ref{fig1}, we illustrate the dependence of the perturbation distribution function $\phi^{-1} (E_{\textbf{k}})$ on energy resulting from inelastic scattering of electrons with LO phonons. We also present the relaxation time for electron-acoustic phonon scattering, $\tau_{ac}^{-1} (E_{\textbf{k}})$, as well as the high-energy relaxation time, $\tau_{HE}^{-1} (E_{\textbf{k}})$, and the low-temperature relaxation time, $\tau_{LT}^{-1} (E_{\textbf{k}})$. Considering both finite and infinite square quantum wells, this data is shown at a density of $N_{s}$ = $10^{11}$ 
 $\mathrm{cm}^{-2}$ for two different well widths and two distinct temperatures From the figure, we observe that the oscillation peaks in the perturbation distribution function $\phi^{-1} (E_{\textbf{k}})$ corresponding to multiples of the energy $\hbar\omega_0$ originate from the phonon emission when the electron energy is equal to $\hbar\omega_0$ and from the coupling of $\phi^{-1} (E_{\textbf{k}})$ with $\phi^{-1} (E_{\textbf{k}}\pm\hbar\omega_0)$. At low carrier temperatures and energies ($E < 9 E_F$) with quantum wells of $V$ = 111.85 meV, and at energies ($E < 6E_F$) with quantum wells of $V$ = 508.25 meV, it is significant to note that the inverse relaxation time $\phi^{-1} (E_{\textbf{k}})$ coincides with $\tau_{ac}^{-1} (E_{\textbf{k}})$ as stated in equation (25), and $\tau_{HE}^{-1} (E_{\textbf{k}})$ as shown in equation (26). This alignment is important and occurs regardless of the quantum well width because, at low temperatures and high energies, the relaxation times are predominantly governed by acoustic phonon scattering. The values of $\phi^{-1} (E_{\textbf{k}})$, $\tau_{ac}^{-1} (E_{\textbf{k}})$, $\tau_{LT}^{-1} (E_{\textbf{k}})$ and $\tau_{HE}^{-1} (E_{\textbf{k}})$ are larger for smaller $V$. Fig. \ref{fig1} also shows as expected that the high energy relaxation time $\tau_{HE}^{-1} (E_{\textbf{k}})$ reaches repeatable results for large $E_{\textbf{k}}$, especially at high temperatures.

\begin{figure*}[!htb] \centering
\includegraphics[width=0.98\textwidth]{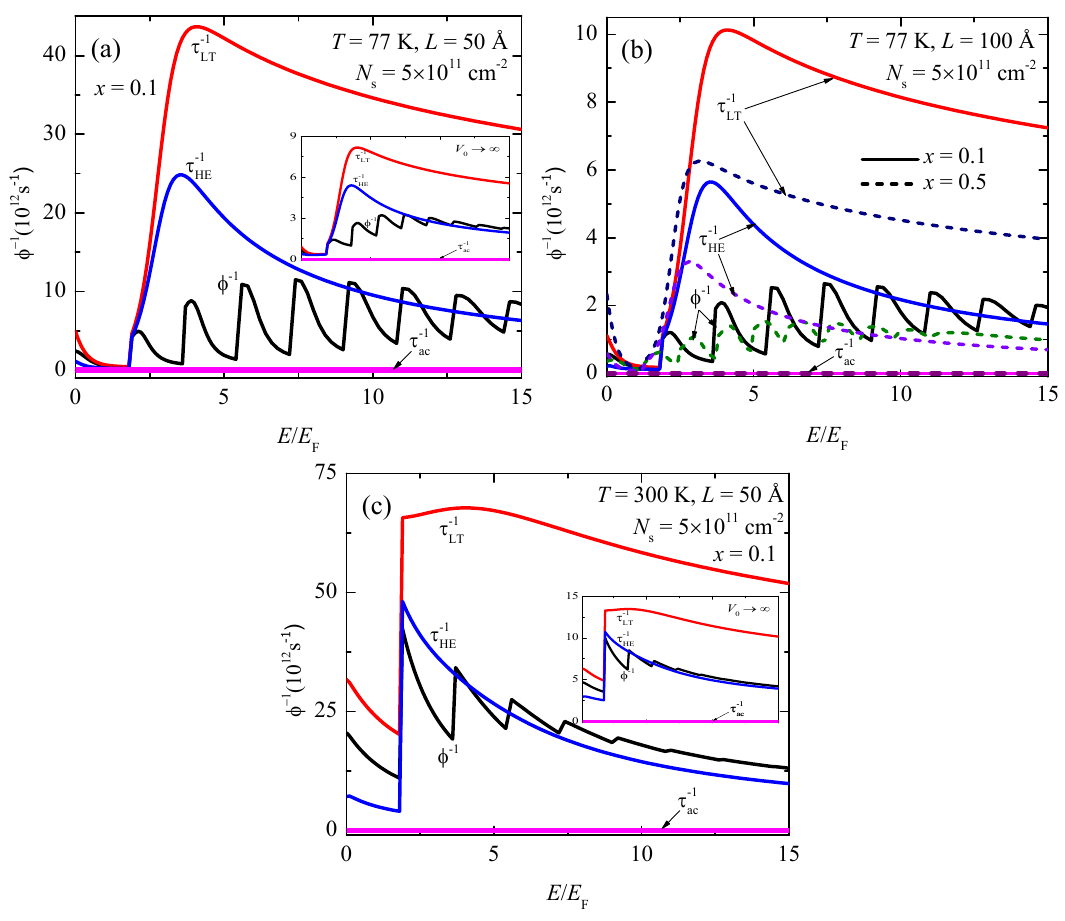}
\caption{Dependence of the perturbation distribution function $\phi^{-1} (E_{\textbf{k}})$ on $E$, the acoustic phonon scattering relaxation time $\tau_{ac}^{-1} (E_{\textbf{k}})$, the high-energy relaxation time $\tau_{HE}^{-1} (E_{\textbf{k}})$, and the low-temperature relaxation time $\tau_{LT}^{-1} (E_{\textbf{k}})$ for finite and infinite square quantum wells of $\text{GaAs}/\text{In}_{x}\text{Ga}_{1-x}\text{As}/\text{GaAs}$ with $N_{s}$ = 5$\times 10^{11}$ 
 $\mathrm{cm}^{-2}$ : (a) For a well width $L$ = 50 $\mathrm{\AA}$ and a temperature $T$ = 77 K with a concentration $x$ = 0.1 and an infinite potential well ($V\to\infty$); (b) For a well width $L$ = 100 $\mathrm{\AA}$ and a temperature $T$ = 77 K for both concentration $x$ = 0.1 and x = 0.5;(c) For a well width $L$ = 50 $\mathrm{\AA}$ and a temperature $T$ = 300 K for a concentration $x$ = 0.1 and an infinite potential well($V\to\infty$).}
 \label{fig2}
\end{figure*}

Figure \ref{fig2} illustrates the dependence of the perturbation distribution function $\phi^{-1} (E_{\textbf{k}})$ on $E$ and the acoustic phonon scattering relaxation time $\tau_{ac}^{-1} (E_{\textbf{k}})$, alongside the high-energy relaxation time $\tau_{HE}^{-1} (E_{\textbf{k}})$ and the low-temperature relaxation time $\tau_{LT}^{-1} (E_{\textbf{k}})$ at a fixed density $N_{s}$ = 5$\times 10^{11}$ 
 $\mathrm{cm}^{-2}$ for two different well widths and temperatures. The data presented in Fig. 1 and Fig. 2 indicate that $\tau_{HE}^{-1} (E_{\textbf{k}})$ achieves repeatable results for large $E_{\textbf{k}}$ under lower density conditions. Notably, the sharpness of the peaks in $\phi^{-1} (E_{\textbf{k}})$ becomes smoother at higher densities, while $\tau_{HE}^{-1} (E_{\textbf{k}})$ closely matches the results derived from the iterative method across the entire carrier energy spectrum. This trend is reinforced by similar findings in both infinite square wells of AlGaN/GaN/AlGaN and finite and infinite triangular quantum wells of AlGaN/GaN \cite{Tai_17},\cite{Tuan_18}. Such insights underscore our results' robustness and relevance in the field.
 
 \begin{figure*}[!h] \centering
\includegraphics[width=0.98\textwidth]{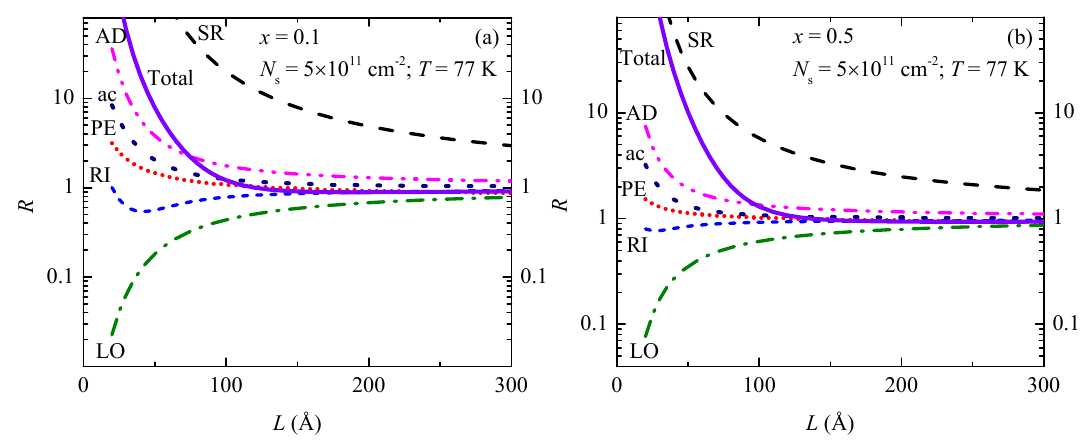}
\caption{The ratio $R =\mu_{fin}/\mu_{inf}$, which represents the mobility of a finite square quantum well relative to that of an infinite quantum well, depends on various scattering mechanisms and the width of the well. The observations are as follows: (a) $N_{s}$ = 5$\times 10^{11}$ 
 $\mathrm{cm}^{-2}$, $T$ = 77 K, $x$ = 0.1; (b) $N_{s}$ = 5$\times 10^{11}$ 
 $\mathrm{cm}^{-2}$, $T$ = 77 K, $x$ = 0.5.}
 \label{fig3}
 \end{figure*}
The ratio $R =\mu_{fin}/\mu_{inf}$ compares the mobility of a finite square quantum well to that of an infinite square quantum well. This ratio varies with the well width for different scattering mechanisms. The $R$ values between $x$ = 0.1 and its corresponding infinite square quantum well, and between $x$ = 0.5 and its corresponding infinite square quantum well, are illustrated in Fig. \ref{fig3}. Importantly, we find that the ratio $R$ for RI and LO scattering remains consistently below 1. This indicates that RI scattering is characterized by long-range Coulomb interactions; while LO scattering is an inelastic process. The observed trends in the ratio $R$ for LO and RI scattering align similarly with those seen in finite and infinite triangular quantum wells \cite{Tai_17}, emphasizing the significance of understanding these relationships in quantum wells. Meanwhile, the R ratio for ac, SR, AD, and PE scattering consistently exceeds 1, with SR scattering demonstrating a particularly significant increase at narrow well widths. At wider well widths, the $R$ ratio stabilizes around 1 for ac, RI, and PE scattering. In conditions of low density and temperature, the total mobility largely depends on SR and ac scattering, which results in $R_\mathrm{total}$ greater than 1 at narrow well widths and approximately equal to 1 at broader well widths. Furthermore, when the barrier height increases, $R_\mathrm{total}$ remains greater than 1 for narrow well widths ($L$ $<$ 100 $\mathrm{\AA}$) because, in this regime, an increase in x allows ac scattering to dominate, as shown in Fig. 6. Consequently, at narrow well widths ($L$ $<$ 100 $\mathrm{\AA}$), $R_\mathrm{total}$ is greater than 1 at low temperatures and densities.	

\begin{figure*}[!h] \centering
\includegraphics[width=0.98\textwidth]{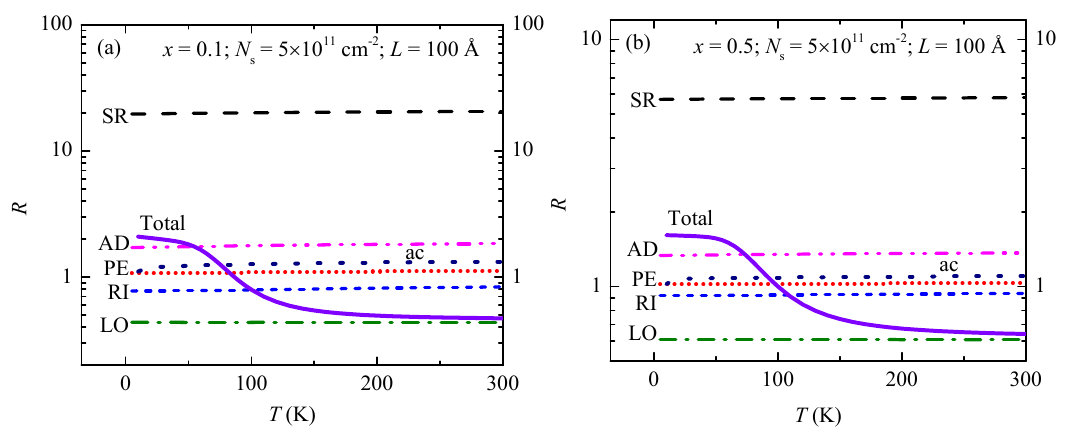}
\caption{The dependence $R =\mu_{fin}/\mu_{inf}$ between finite and infinite square quantum well mobility for different scattering mechanisms with temperature: (a) $N_{s}$ = 5$\times 10^{11}$ 
 $\mathrm{cm}^{-2}$, $L$ = 100 $\mathrm{\AA}$, $x$ = 0.1; (b) $N_{s}$ = 5$\times 10^{11}$ 
 $\mathrm{cm}^{-2}$, $L$ = 100 $\mathrm{\AA}$, $x$ = 0.5.}
 \label{fig4}
\end{figure*}

Figure \ref{fig4} illustrates how the $R$ ratio varies with temperature for different scattering mechanisms at $N_{s}$ = 5$\times 10^{11}$ $\mathrm{cm}^{-2}$, $L$ =  100  $\mathrm{\AA}$, corresponding to two values of finite quantum well $x$ = 0.1 and $x$ = 0.5. We observe that the scattering ratios tend to stabilize with temperature. Specifically, the ratios $R_\mathrm{RI}$ and $R_\mathrm{LO}$ are less than 1, while $R_\mathrm{AD}$, $R_\mathrm{ac}$, $R_\mathrm{PE}$, and $R_\mathrm{SR}$ are greater than 1, with $R_\mathrm{PE}$ being approximately equal to 1. At low temperatures ($T$ $<$ 77 K), the total ratio $R_\mathrm{total}$ is consistently greater than 1 and decreases with temperature, reaching its lowest value at $T$ = 300 K. This decrease occurs because, at this temperature, LO scattering ($R_\mathrm{LO} < 1$) is dominant.

\begin{figure*}[!h] \centering
\includegraphics[width=0.98\textwidth]{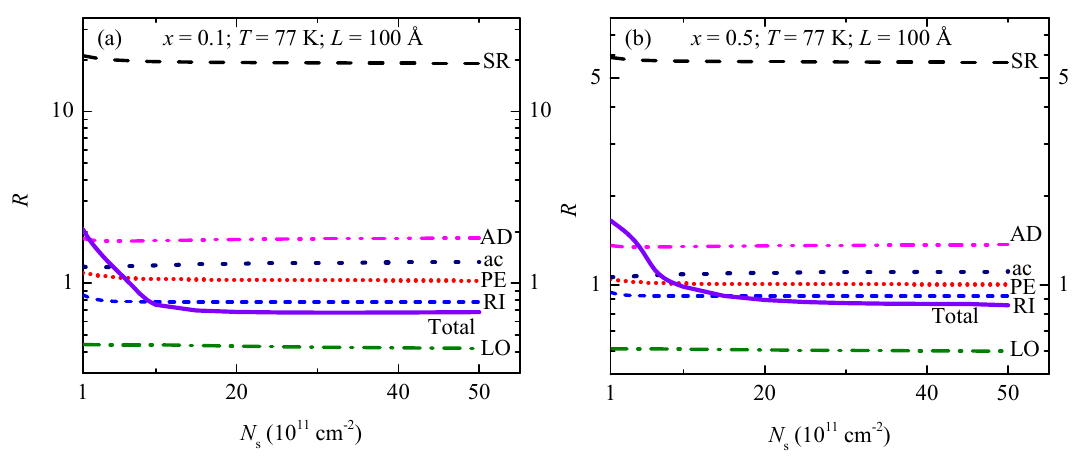}
\caption{The dependence $R$ is defined as the ratio of the mobility of finite QWs ($\mu_{fin}$) to that of infinite quantum wells ($\mu_{inf}$) for various scattering mechanisms in relation to density. The conditions are as follows: (a) at a temperature of 77 K with well width $L$ = 100 $\mathrm{\AA}$, and $x$ = 0.1; (b) at a temperature of 77 K with well width $L$ =  100  $\mathrm{\AA}$, and $x$ = 0.5.}
 \label{fig5}
\end{figure*}
Figure \ref{fig5} shows the variation of the $R$ ratio with density for $T$ = 77 K, $L$ =  100  $\mathrm{\AA}$, corresponding to two values of finite quantum well $x$ = 0.1 and $x$ = 0.5. From the figure, we observe that the R ratio is almost saturated with density for scattering mechanisms. Notably, the $R_\mathrm{total}$ is greater than 1 at low density, temperature and well width. At higher densities $N_s$ $>$ 5$\times 10^{11}$ $\mathrm{cm}^{-2}$, $R_\mathrm{total}$ decreases to less than 1 and almost saturates at high densities $N_s$ $>$ 3$\times 10^{12}$ $\mathrm{cm}^{-2}$. More specifically, at $N_s$ $=$ 3$\times 10^{12}$ $\mathrm{cm}^{-2}$ to $N_s$ $=$ 5$\times 10^{12}$ $\mathrm{cm}^{-2}$ for $x$ = 0.1 the $R_\mathrm{total}$ ratio is approximately 0.67, and at $x$ = 0.5 the $R_\mathrm{total}$ ratio is approximately 0.86.

\begin{figure*}
 \centering
 \includegraphics[width=0.985\textwidth]{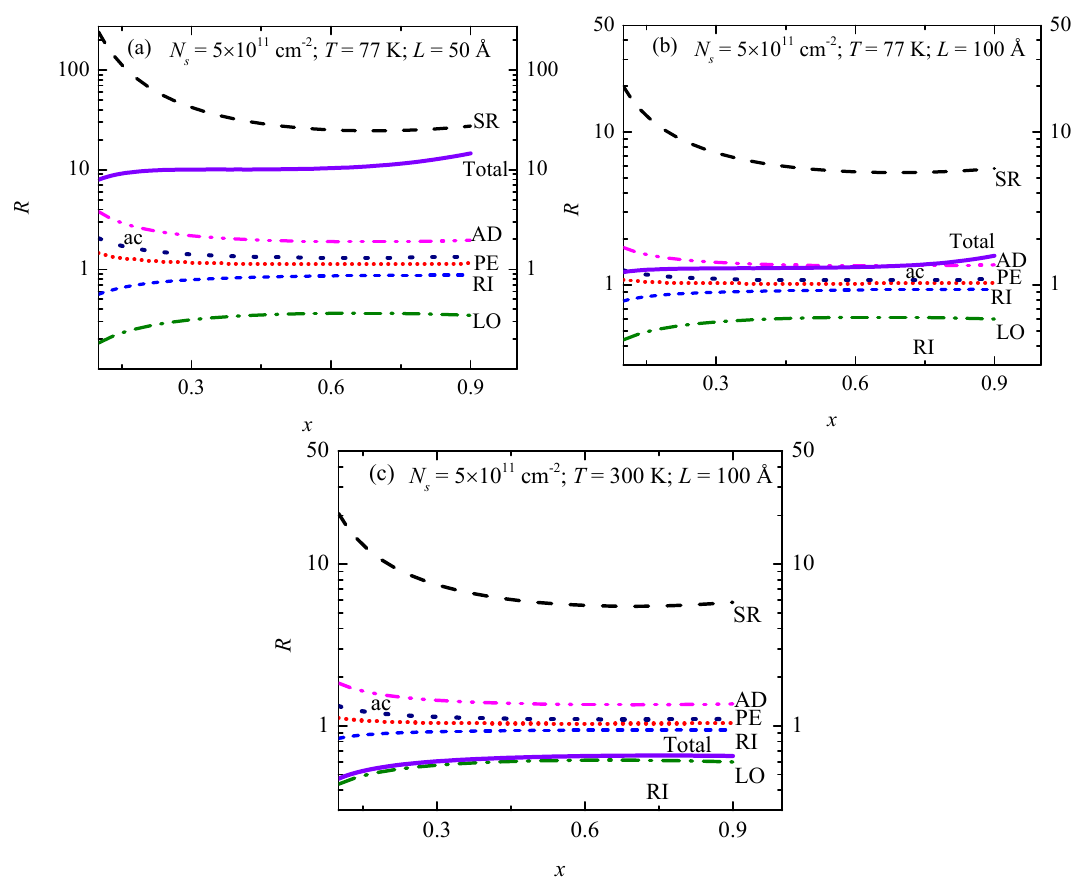}
\caption{Variation of the ratio $R =\mu_{fin}/\mu_{inf}$, the mobility between finite and infinite QWs corresponding to $x$ varying from 0.1 to 0.9 at density $N_s$ $=$ 5$\times 10^{11}$ $\mathrm{cm}^{-2}$: (a) $T$ = 77 K, $L$ = 50 $\mathrm{\AA}$; (b) $T$ = 77 K, $L$ = 100 $\mathrm{\AA}$; and (c) $T$ = 300 K, $L$ = 100 $\mathrm{\AA}$.}
 \label{fig6}
\end{figure*}

Figure \ref{fig6} shows the ratio $R =\mu_{fin}/\mu_{inf}$ between a finite and an infinite square quantum well, respectively, according to the concentration $x$ at the density $N_s$ $=$ 5$\times 10^{11}$ $\mathrm{cm}^{-2}$, for two values of temperature and well width. From the figure, we see that the mobility of the finite quantum well is always larger than that of the infinite quantum well at low temperatures, especially at a narrow well width $L$ = 50 $\mathrm{\AA}$, the ratio is approximately 10. More specifically, with $x$ = 0.9 at $T$ = 77 K, $L$ = 50 $\mathrm{\AA}$: $\mu_{fin}$ $\approx$ 30717 ($\mathrm{cm}^{-2}$/Vs), $\mu_{inf}$ $\approx$ 3250 ($\mathrm{cm}^{-2}$/Vs). At $T$ = 300 K, the ratio $R$ is less than 1, and $R$ is approximately 0.6. In more detail, with $x$ = 0.9 at $T$ = 300 K, $L$ = 100 $\mathrm{\AA}$: $\mu_{fin}$ $\approx$ 6665 ($\mathrm{cm}^{-2}$/Vs), $\mu_{inf}$ $\approx$ 10237 ($\mathrm{cm}^{-2}$/Vs). In addition, comparing the total mobility between finite quantum wells: $N_s$ $=$ 5$\times 10^{11}$ $\mathrm{cm}^{-2}$, $T$ = 77 K, $L$ = 50 $\mathrm{\AA}$, when $x$ = 0.1 then $\mu^\mathrm{0.1}$ $\approx$ 11574 ($\mathrm{cm}^{-2}$/Vs), when $x$ = 0.9 then the mobility increases to $\mu^\mathrm{0.1}$ $\approx$ 30717 ($\mathrm{cm}^{-2}$/Vs). Furthermore, a compelling comparison of total mobility in infinite quantum wells reveals significant insights: $N_s$ $=$ 5$\times 10^{11}$ $\mathrm{cm}^{-2}$, $T$ = 77 K, $L$ = 50 $\mathrm{\AA}$, when $x$ = 0.1 then $\mu^\mathrm{0.1}$ $\approx$ 4749 ($\mathrm{cm}^{-2}$/Vs), when $x$ = 0.9 then the mobility decreases to $\mu^\mathrm{0.9}$ $\approx$ 3250 ($\mathrm{cm}^{-2}$/Vs). In summary, at low density, the mobility of the finite square quantum well - which improves with larger values of $x$-dramatically surpasses that of the infinite square quantum well, where mobility is better for smaller values of $x$. This distinction becomes particularly pronounced at low temperatures and narrow quantum well widths. However, even at room temperature, it's striking to note that the mobility of the finite square quantum well is almost halved compared to its infinite counterpart. This highlights the significant advantages of the finite square quantum well in specific conditions.

 \begin{figure*}  \centering
 \includegraphics[width=0.985\textwidth]{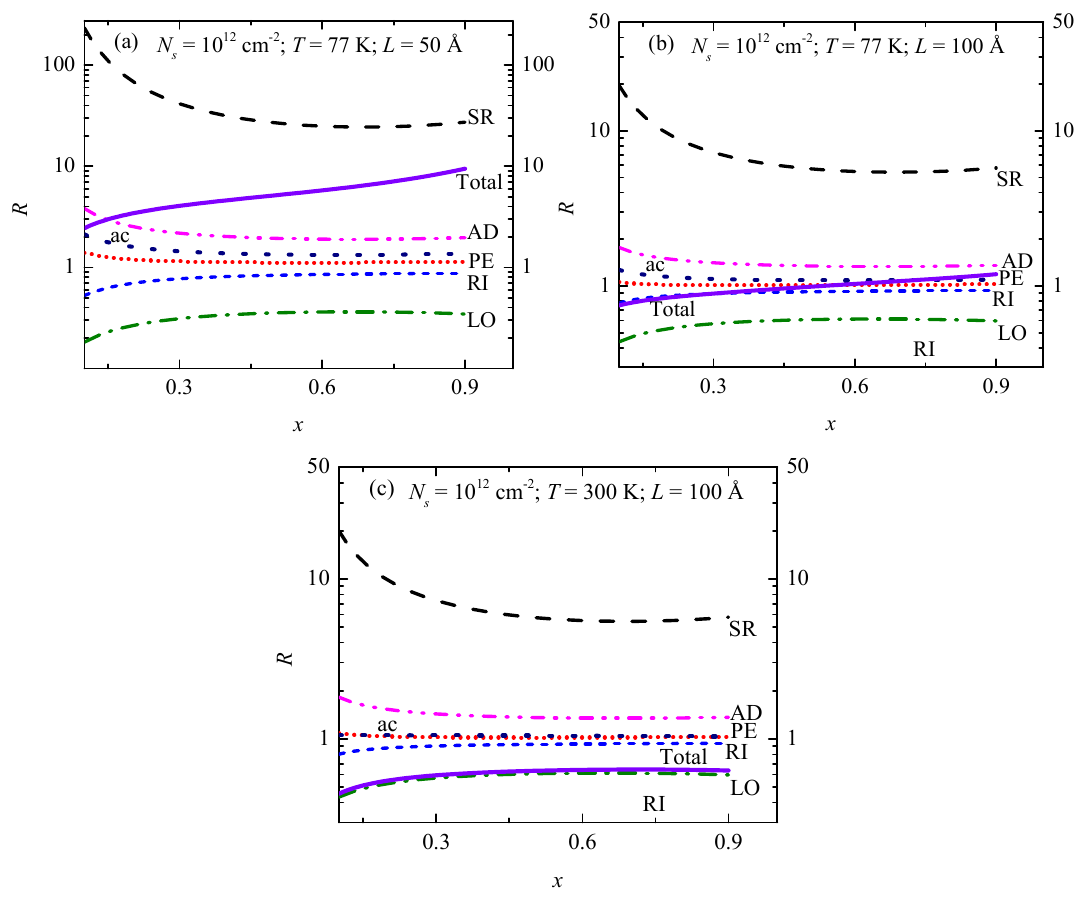}
 \caption{The ratio $R =\mu_{fin}/\mu_{inf}$ illustrates the crucial variation in mobility between finite and infinite quantum wells, as the parameter $x$ shifts from 0.1 to 0.9 at a fixed density of $N_s$ $=$ 1$\times 10^{12}$ $\mathrm{cm}^{-2}$. This analysis spans three distinct conditions: (a) At a low temperature of $T$ = 77 K and a well width of $L$ = 50 $\mathrm{\AA}$; (b) Maintaining the same temperature of $T$ = 77 K but with an expanded well width of $L$ = 100 $\mathrm{\AA}$; (c) At higher temperature of $T$ = 300 K with a well width also set at $L$ = 100 $\mathrm{\AA}$.}
 \label{fig7}
 \end{figure*}

In Fig. \ref{fig7}, we present a compelling analysis of the ratio $R =\mu_{fin}/\mu_{inf}$, which compares the mobilities of finite and infinite square quantum wells as a function of concentration x at a density of $N_s$ $=$ 1$\times 10^{12}$ $\mathrm{cm}^{-2}$. This analysis is conducted over two distinct temperatures and well widths, revealing important insights. The data shows that at low temperatures and narrow well widths, the ratio $R$ demonstrates a gradual increase, indicating a significant relationship between these variables. Notably, when we maintain a constant temperature and increase the well width, we find that R exceeds 1 only at larger concentrations, specifically when $x$ = 0.6. At this concentration, we observe $\mu_{fin}$ $\approx$ 54391 ($\mathrm{cm}^{-2}$/Vs) and $\mu_{inf}$ $\approx$ 52663 ($\mathrm{cm}^{-2}$/Vs). At a temperature of $T$ = 300 K, the ratio $R$ dips below 1, which mirrors the findings at a density of $N_s$ $=$ 5$\times  10^{11}$ $\mathrm{cm}^{-2}$ shown in Figure 6, with the ratio hovering around 0.6. This prompts a powerful conclusion: combining insights from Figures 6 and 7 reinforces the idea that, at 300 K, the mobility of the small finite square quantum well is approximately 0.6 times less than that of the infinite square quantum well at elevated $x$ values. This relationship appears remarkably consistent, demonstrating little dependence on density or well width. To illustrate this further, consider the specific scenarios at $N_s$ $=$ 1$\times 10^{12}$ $\mathrm{cm}^{-2}$, $T$ = 77 K, $L$ = 50 $\mathrm{\AA}$: For $x$ = 0.1, $\mu_{fin}$ $\approx$ 1567 ($\mathrm{cm}^{-2}$/Vs), $\mu_{inf}$ $\approx$ 3449 ($\mathrm{cm}^{-2}$/Vs); For $x$ = 0.9, $\mu_{fin}$ $\approx$ 7171 ($\mathrm{cm}^{-2}$/Vs), $\mu_{inf}$ $\approx$ 11299 ($\mathrm{cm}^{-2}$/Vs). These findings not only highlight the critical differences in mobility between finite and infinite quantum wells but also underscore the significant implications for the understanding of quantum mobility in various conditions.
 
\section{Conclusions}
The geometry of quantum wells (QWs) fundamentally dictates carrier mobility, yet the distinct advantages of square QWs over triangular or infinite counterparts have remained underexplored. In this study, we systematically compared carrier mobility in finite and infinite square QWs based on GaAs/In$_x$Ga$_{1-x}$As/GaAs heterostructures, incorporating critical scattering mechanisms such as remote impurities (RI), alloy disorder (AD), and longitudinally polarized optical (LO) phonons. Our findings demonstrate that finite square QWs achieve superior mobility ($R_{total}>$ 1 at low temperatures ($T$ = 77 K), narrow well widths ($L <$ 100\AA), and low carrier densities, driven by suppressed RI scattering and enhanced AD scattering. In contrast, infinite QWs dominate at room temperature (
$T$ =  300 K), where LO phonon scattering ($R_{LO}<$ 1 governs transport—a trend starkly different from triangular QWs, where asymmetric confinement alters scattering dynamics \cite{Tai_17}.

Crucially, the symmetric confinement in square QWs enables precise tuning of subband wavefunctions via indium content ($x$) and well width ($L$), offering unparalleled flexibility in device design. For instance, at $x$ = 0.9, finite QWs achieve mobilities exceeding 30,000 cm$^2$/Vs under low-temperature conditions, far surpassing infinite QWs. This tunability contrasts with triangular QWs, where mobility trends are constrained by geometric asymmetry \cite{Tai_17}. At room temperature, however, infinite QWs outperform by up to 40\%, underscoring the necessity of geometry-specific optimization.

These insights are transformative for semiconductor device engineering. For low-temperature applications like quantum computing or cryogenic HEMTs, finite square QWs with high $x$ are optimal. Conversely, infinite QWs are preferable for high-temperature optoelectronics or power devices. By bridging the gap between theoretical models and practical design principles, this work empowers manufacturers to tailor QW geometries for targeted performance, advancing the frontier of III-V semiconductor technology.

\bibliographystyle{aipnum4-1}

\bibliography{manu_SSC}

\begin{thebibliography}{20}%
\makeatletter
\providecommand \@ifxundefined [1]{%
 \@ifx{#1\undefined}
}%
\providecommand \@ifnum [1]{%
 \ifnum #1\expandafter \@firstoftwo
 \else \expandafter \@secondoftwo
 \fi
}%
\providecommand \@ifx [1]{%
 \ifx #1\expandafter \@firstoftwo
 \else \expandafter \@secondoftwo
 \fi
}%
\providecommand \natexlab [1]{#1}%
\providecommand \enquote  [1]{``#1''}%
\providecommand \bibnamefont  [1]{#1}%
\providecommand \bibfnamefont [1]{#1}%
\providecommand \citenamefont [1]{#1}%
\providecommand \href@noop [0]{\@secondoftwo}%
\providecommand \href [0]{\begingroup \@sanitize@url \@href}%
\providecommand \@href[1]{\@@startlink{#1}\@@href}%
\providecommand \@@href[1]{\endgroup#1\@@endlink}%
\providecommand \@sanitize@url [0]{\catcode `\\12\catcode `\$12\catcode `\&12\catcode `\#12\catcode `\^12\catcode `\_12\catcode `\%12\relax}%
\providecommand \@@startlink[1]{}%
\providecommand \@@endlink[0]{}%
\providecommand \url  [0]{\begingroup\@sanitize@url \@url }%
\providecommand \@url [1]{\endgroup\@href {#1}{\urlprefix }}%
\providecommand \urlprefix  [0]{URL }%
\providecommand \Eprint [0]{\href }%
\providecommand \doibase [0]{https://doi.org/}%
\providecommand \selectlanguage [0]{\@gobble}%
\providecommand \bibinfo  [0]{\@secondoftwo}%
\providecommand \bibfield  [0]{\@secondoftwo}%
\providecommand \translation [1]{[#1]}%
\providecommand \BibitemOpen [0]{}%
\providecommand \bibitemStop [0]{}%
\providecommand \bibitemNoStop [0]{.\EOS\space}%
\providecommand \EOS [0]{\spacefactor3000\relax}%
\providecommand \BibitemShut  [1]{\csname bibitem#1\endcsname}%
\let\auto@bib@innerbib\@empty
\bibitem [{\citenamefont {Arent}\ \emph {et~al.}(1989)\citenamefont {Arent}, \citenamefont {Deneffe}, \citenamefont {Van~Hoof}, \citenamefont {De~Boeck},\ and\ \citenamefont {Borghs}}]{Arent}%
  \BibitemOpen
  \bibfield  {author} {\bibinfo {author} {\bibfnamefont {D.~J.}\ \bibnamefont {Arent}}, \bibinfo {author} {\bibfnamefont {K.}~\bibnamefont {Deneffe}}, \bibinfo {author} {\bibfnamefont {C.}~\bibnamefont {Van~Hoof}}, \bibinfo {author} {\bibfnamefont {J.}~\bibnamefont {De~Boeck}},\ and\ \bibinfo {author} {\bibfnamefont {G.}~\bibnamefont {Borghs}},\ }\href {https://doi.org/10.1063/1.344395} {\bibfield  {journal} {\bibinfo  {journal} {J. Appl. Phys.}\ }\textbf {\bibinfo {volume} {66}},\ \bibinfo {pages} {1739} (\bibinfo {year} {1989})}\BibitemShut {NoStop}%
\bibitem [{\citenamefont {Bedoui}\ \emph {et~al.}(2016)\citenamefont {Bedoui}, \citenamefont {Habchi}, \citenamefont {Moussa}, \citenamefont {Rebey},\ and\ \citenamefont {El~Jani}}]{Bedoui}%
  \BibitemOpen
  \bibfield  {author} {\bibinfo {author} {\bibfnamefont {M.}~\bibnamefont {Bedoui}}, \bibinfo {author} {\bibfnamefont {M.}~\bibnamefont {Habchi}}, \bibinfo {author} {\bibfnamefont {I.}~\bibnamefont {Moussa}}, \bibinfo {author} {\bibfnamefont {A.}~\bibnamefont {Rebey}},\ and\ \bibinfo {author} {\bibfnamefont {B.}~\bibnamefont {El~Jani}},\ }\href {https://doi.org/10.1016/j.surfcoat.2015.10.002} {\bibfield  {journal} {\bibinfo  {journal} {Surf. Coat. Tech.}\ }\textbf {\bibinfo {volume} {295}},\ \bibinfo {pages} {107} (\bibinfo {year} {2016})}\BibitemShut {NoStop}%
\bibitem [{\citenamefont {Vy}\ \emph {et~al.}(2020)\citenamefont {Vy}, \citenamefont {Minh}, \citenamefont {Tuyet~Anh}, \citenamefont {Trien},\ and\ \citenamefont {Hien}}]{Vy2020superlattice}%
  \BibitemOpen
  \bibfield  {author} {\bibinfo {author} {\bibfnamefont {N.~D.}\ \bibnamefont {Vy}}, \bibinfo {author} {\bibfnamefont {L.~N.}\ \bibnamefont {Minh}}, \bibinfo {author} {\bibfnamefont {N.~T.}\ \bibnamefont {Tuyet~Anh}}, \bibinfo {author} {\bibfnamefont {H.~D.}\ \bibnamefont {Trien}},\ and\ \bibinfo {author} {\bibfnamefont {N.~D.}\ \bibnamefont {Hien}},\ }\href {https://doi.org/10.1016/j.spmi.2020.106626} {\bibfield  {journal} {\bibinfo  {journal} {Superlatt. Microstruct.}\ }\textbf {\bibinfo {volume} {145}},\ \bibinfo {pages} {106626} (\bibinfo {year} {2020})}\BibitemShut {NoStop}%
\bibitem [{\citenamefont {Vainberg}\ \emph {et~al.}(2013)\citenamefont {Vainberg}, \citenamefont {Pylypchuk}, \citenamefont {Baidus},\ and\ \citenamefont {Zvonkov}}]{Vainberg2013}%
  \BibitemOpen
  \bibfield  {author} {\bibinfo {author} {\bibfnamefont {V.}~\bibnamefont {Vainberg}}, \bibinfo {author} {\bibfnamefont {A.}~\bibnamefont {Pylypchuk}}, \bibinfo {author} {\bibfnamefont {N.}~\bibnamefont {Baidus}},\ and\ \bibinfo {author} {\bibfnamefont {B.}~\bibnamefont {Zvonkov}},\ }\href {https://core.ac.uk/download/pdf/87401892.pdf} {\bibfield  {journal} {\bibinfo  {journal} {Semicon. Phys. Quan. Elec. \& Optoelec.}\ }\textbf {\bibinfo {volume} {16}},\ \bibinfo {pages} {152} (\bibinfo {year} {2013})}\BibitemShut {NoStop}%
\bibitem [{\citenamefont {Quang}, \citenamefont {Tuoc},\ and\ \citenamefont {Huan}(2003)}]{Quang_13}%
  \BibitemOpen
  \bibfield  {author} {\bibinfo {author} {\bibfnamefont {D.~N.}\ \bibnamefont {Quang}}, \bibinfo {author} {\bibfnamefont {V.~N.}\ \bibnamefont {Tuoc}},\ and\ \bibinfo {author} {\bibfnamefont {T.~D.}\ \bibnamefont {Huan}},\ }\href {https://doi.org/10.1103/PhysRevB.68.195316} {\bibfield  {journal} {\bibinfo  {journal} {Phys. Rev. B}\ }\textbf {\bibinfo {volume} {68}},\ \bibinfo {pages} {195316} (\bibinfo {year} {2003})}\BibitemShut {NoStop}%
\bibitem [{\citenamefont {Truong}\ \emph {et~al.}(2020)\citenamefont {Truong}, \citenamefont {Nguyen}, \citenamefont {Vo},\ and\ \citenamefont {Dang}}]{Tuan_16}%
  \BibitemOpen
  \bibfield  {author} {\bibinfo {author} {\bibfnamefont {V.~T.}\ \bibnamefont {Truong}}, \bibinfo {author} {\bibfnamefont {Q.~K.}\ \bibnamefont {Nguyen}}, \bibinfo {author} {\bibfnamefont {V.~T.}\ \bibnamefont {Vo}},\ and\ \bibinfo {author} {\bibfnamefont {K.~L.}\ \bibnamefont {Dang}},\ }\href {https://doi.org/10.15625/0868-3166/30/2/14446} {\bibfield  {journal} {\bibinfo  {journal} {Commun. in Phys.}\ }\textbf {\bibinfo {volume} {30}},\ \bibinfo {pages} {123} (\bibinfo {year} {2020})}\BibitemShut {NoStop}%
\bibitem [{\citenamefont {Nestoklon}\ \emph {et~al.}(2016)\citenamefont {Nestoklon}, \citenamefont {Tarasenko}, \citenamefont {Benchamekh},\ and\ \citenamefont {Voisin}}]{Nestoklon_16}%
  \BibitemOpen
  \bibfield  {author} {\bibinfo {author} {\bibfnamefont {M.~O.}\ \bibnamefont {Nestoklon}}, \bibinfo {author} {\bibfnamefont {S.~A.}\ \bibnamefont {Tarasenko}}, \bibinfo {author} {\bibfnamefont {R.}~\bibnamefont {Benchamekh}},\ and\ \bibinfo {author} {\bibfnamefont {P.}~\bibnamefont {Voisin}},\ }\href {https://doi.org/10.1103/PhysRevB.94.115310} {\bibfield  {journal} {\bibinfo  {journal} {Phys. Rev. B}\ }\textbf {\bibinfo {volume} {94}},\ \bibinfo {pages} {115310} (\bibinfo {year} {2016})}\BibitemShut {NoStop}%
\bibitem [{\citenamefont {Kawamura}\ and\ \citenamefont {Das~Sarma}(1992)}]{Kawamura}%
  \BibitemOpen
  \bibfield  {author} {\bibinfo {author} {\bibfnamefont {T.}~\bibnamefont {Kawamura}}\ and\ \bibinfo {author} {\bibfnamefont {S.}~\bibnamefont {Das~Sarma}},\ }\href {https://doi.org/10.1103/PhysRevB.45.3612} {\bibfield  {journal} {\bibinfo  {journal} {Phys. Rev. B}\ }\textbf {\bibinfo {volume} {45}},\ \bibinfo {pages} {3612} (\bibinfo {year} {1992})}\BibitemShut {NoStop}%
\bibitem [{\citenamefont {Van~Tai}, \citenamefont {Van~Tuan},\ and\ \citenamefont {Vy}(2025)}]{Tai_17}%
  \BibitemOpen
  \bibfield  {author} {\bibinfo {author} {\bibfnamefont {V.}~\bibnamefont {Van~Tai}}, \bibinfo {author} {\bibfnamefont {T.}~\bibnamefont {Van~Tuan}},\ and\ \bibinfo {author} {\bibfnamefont {N.~D.}\ \bibnamefont {Vy}},\ }\href {https://doi.org/10.1007/s11664-025-11954-z} {\bibfield  {journal} {\bibinfo  {journal} {J. Electron. Mater.}\ } (\bibinfo {year} {2025}),\ 10.1007/s11664-025-11954-z}\BibitemShut {NoStop}%
\bibitem [{\citenamefont {Hamaguchi}\ and\ \citenamefont {Hamaguchi}(2010)}]{Hamaguchi}%
  \BibitemOpen
  \bibfield  {author} {\bibinfo {author} {\bibfnamefont {C.}~\bibnamefont {Hamaguchi}}\ and\ \bibinfo {author} {\bibfnamefont {C.}~\bibnamefont {Hamaguchi}},\ }\href@noop {} {\emph {\bibinfo {title} {Basic semiconductor physics}}},\ Vol.~\bibinfo {volume} {9}\ (\bibinfo  {publisher} {Springer},\ \bibinfo {year} {2010})\BibitemShut {NoStop}%
\bibitem [{\citenamefont {Van~Tuan}\ \emph {et~al.}(2021)\citenamefont {Van~Tuan}, \citenamefont {Khanh}, \citenamefont {Van~Tai},\ and\ \citenamefont {Linh}}]{Tuan_18}%
  \BibitemOpen
  \bibfield  {author} {\bibinfo {author} {\bibfnamefont {T.}~\bibnamefont {Van~Tuan}}, \bibinfo {author} {\bibfnamefont {N.~Q.}\ \bibnamefont {Khanh}}, \bibinfo {author} {\bibfnamefont {V.}~\bibnamefont {Van~Tai}},\ and\ \bibinfo {author} {\bibfnamefont {D.~K.}\ \bibnamefont {Linh}},\ }\href {https://doi.org/10.1140/epjb/s10051-021-00111-0} {\bibfield  {journal} {\bibinfo  {journal} {Eur. Phys. J. B}\ }\textbf {\bibinfo {volume} {94}} (\bibinfo {year} {2021}),\ 10.1140/epjb/s10051-021-00111-0}\BibitemShut {NoStop}%
\bibitem [{\citenamefont {Khanh}(2011)}]{Khanh_22}%
  \BibitemOpen
  \bibfield  {author} {\bibinfo {author} {\bibfnamefont {N.~Q.}\ \bibnamefont {Khanh}},\ }\href {https://doi.org/10.1016/j.physe.2011.05.028} {\bibfield  {journal} {\bibinfo  {journal} {Phys. E: Low-dim. Sys. Nanostruct.}\ }\textbf {\bibinfo {volume} {43}},\ \bibinfo {pages} {1712–1716} (\bibinfo {year} {2011})}\BibitemShut {NoStop}%
\bibitem [{\citenamefont {Khanh}\ and\ \citenamefont {Tai}(2014)}]{Khanh_23}%
  \BibitemOpen
  \bibfield  {author} {\bibinfo {author} {\bibfnamefont {N.~Q.}\ \bibnamefont {Khanh}}\ and\ \bibinfo {author} {\bibfnamefont {V.~V.}\ \bibnamefont {Tai}},\ }\href {https://doi.org/10.1016/j.physe.2013.11.021} {\bibfield  {journal} {\bibinfo  {journal} {Phys. E: Low-dim. Sys. Nanostruct.}\ }\textbf {\bibinfo {volume} {58}},\ \bibinfo {pages} {84–87} (\bibinfo {year} {2014})}\BibitemShut {NoStop}%
\bibitem [{\citenamefont {Tai}\ and\ \citenamefont {Khanh}(2015)}]{Tai_24}%
  \BibitemOpen
  \bibfield  {author} {\bibinfo {author} {\bibfnamefont {V.~V.}\ \bibnamefont {Tai}}\ and\ \bibinfo {author} {\bibfnamefont {N.~Q.}\ \bibnamefont {Khanh}},\ }\href {https://doi.org/10.1016/j.physe.2014.11.015} {\bibfield  {journal} {\bibinfo  {journal} {Phys. E: Low-dim. Sys. Nanostruct.}\ }\textbf {\bibinfo {volume} {67}},\ \bibinfo {pages} {84–88} (\bibinfo {year} {2015})}\BibitemShut {NoStop}%
\bibitem [{\citenamefont {Van~Tuan}\ \emph {et~al.}(2023)\citenamefont {Van~Tuan}, \citenamefont {Khanh}, \citenamefont {Van~Tai},\ and\ \citenamefont {Linh}}]{Tuan_25}%
  \BibitemOpen
  \bibfield  {author} {\bibinfo {author} {\bibfnamefont {T.}~\bibnamefont {Van~Tuan}}, \bibinfo {author} {\bibfnamefont {N.~Q.}\ \bibnamefont {Khanh}}, \bibinfo {author} {\bibfnamefont {V.}~\bibnamefont {Van~Tai}},\ and\ \bibinfo {author} {\bibfnamefont {D.~K.}\ \bibnamefont {Linh}},\ }\href {https://doi.org/10.1007/s12648-023-02662-7} {\bibfield  {journal} {\bibinfo  {journal} {Indian J. Phys.}\ }\textbf {\bibinfo {volume} {97}},\ \bibinfo {pages} {2961–2969} (\bibinfo {year} {2023})}\BibitemShut {NoStop}%
\bibitem [{\citenamefont {Quang}\ \emph {et~al.}(2007)\citenamefont {Quang}, \citenamefont {Tung}, \citenamefont {Hien},\ and\ \citenamefont {Huy}}]{Quang_26}%
  \BibitemOpen
  \bibfield  {author} {\bibinfo {author} {\bibfnamefont {D.~N.}\ \bibnamefont {Quang}}, \bibinfo {author} {\bibfnamefont {N.~H.}\ \bibnamefont {Tung}}, \bibinfo {author} {\bibfnamefont {D.~T.}\ \bibnamefont {Hien}},\ and\ \bibinfo {author} {\bibfnamefont {H.~A.}\ \bibnamefont {Huy}},\ }\href {https://doi.org/10.1103/PhysRevB.75.073305} {\bibfield  {journal} {\bibinfo  {journal} {Phys. Rev. B}\ }\textbf {\bibinfo {volume} {75}},\ \bibinfo {pages} {073305} (\bibinfo {year} {2007})}\BibitemShut {NoStop}%
\bibitem [{\citenamefont {Rode}(1970)}]{Rode}%
  \BibitemOpen
  \bibfield  {author} {\bibinfo {author} {\bibfnamefont {D.~L.}\ \bibnamefont {Rode}},\ }\href {https://doi.org/10.1103/PhysRevB.2.1012} {\bibfield  {journal} {\bibinfo  {journal} {Phys. Rev. B}\ }\textbf {\bibinfo {volume} {2}},\ \bibinfo {pages} {1012} (\bibinfo {year} {1970})}\BibitemShut {NoStop}%
\bibitem [{\citenamefont {Fraj}\ \emph {et~al.}(2016)\citenamefont {Fraj}, \citenamefont {Saidi}, \citenamefont {Bouzaiene}, \citenamefont {Sfaxi},\ and\ \citenamefont {Maaref}}]{Fraj}%
  \BibitemOpen
  \bibfield  {author} {\bibinfo {author} {\bibfnamefont {I.}~\bibnamefont {Fraj}}, \bibinfo {author} {\bibfnamefont {F.}~\bibnamefont {Saidi}}, \bibinfo {author} {\bibfnamefont {L.}~\bibnamefont {Bouzaiene}}, \bibinfo {author} {\bibfnamefont {L.}~\bibnamefont {Sfaxi}},\ and\ \bibinfo {author} {\bibfnamefont {H.}~\bibnamefont {Maaref}},\ }\href {https://doi.org/10.1016/j.optmat.2016.05.021} {\bibfield  {journal} {\bibinfo  {journal} {Opt. Mater.}\ }\textbf {\bibinfo {volume} {58}},\ \bibinfo {pages} {121–127} (\bibinfo {year} {2016})}\BibitemShut {NoStop}%
\bibitem [{\citenamefont {Quay}(2001)}]{Quay}%
  \BibitemOpen
  \bibfield  {author} {\bibinfo {author} {\bibfnamefont {R.}~\bibnamefont {Quay}},\ }\emph {\bibinfo {title} {Analysis and simulation of high electron mobility transistors}},\ \href {https://doi.org/10.34726/HSS.2001.03341873} {Ph.D. thesis} (\bibinfo {year} {2001})\BibitemShut {NoStop}%
\bibitem [{\citenamefont {Van-Tan}\ \emph {et~al.}(2019)\citenamefont {Van-Tan}, \citenamefont {Thang}, \citenamefont {Vy},\ and\ \citenamefont {Cao}}]{VanTan19PLA}%
  \BibitemOpen
  \bibfield  {author} {\bibinfo {author} {\bibfnamefont {L.}~\bibnamefont {Van-Tan}}, \bibinfo {author} {\bibfnamefont {T.~V.}\ \bibnamefont {Thang}}, \bibinfo {author} {\bibfnamefont {N.~D.}\ \bibnamefont {Vy}},\ and\ \bibinfo {author} {\bibfnamefont {H.~T.}\ \bibnamefont {Cao}},\ }\href {https://doi.org/10.1016/j.physleta.2019.04.004} {\bibfield  {journal} {\bibinfo  {journal} {Phys. Lett. A}\ }\textbf {\bibinfo {volume} {383}},\ \bibinfo {pages} {2110–2113} (\bibinfo {year} {2019})}\BibitemShut {NoStop}%
\end{thebibliography}%

\end{document}